\documentclass[5p]{elsarticle}

\usepackage{amsmath}
\usepackage{graphicx}
\usepackage[fixpdftex]{xcolor}
\usepackage{pdfcolmk}
\usepackage{natbib}
\biboptions{square,numbers,sort&compress}

\newcommand{\vr}{\vv{r}}
\newcommand{\piv}{\Pi_\text{v}}
\newcommand{\pith}{\Pi_\text{th}}
\newcommand{\pivet}{\Pi_\text{v}^*}
\newcommand{\pithet}{\Pi_\text{th}^*}
\newcommand{\pivetlin}{\Pi_\text{v,lin}^*}
\newcommand{\pithetlin}{\Pi_\text{th,lin}^*}
\newcommand{\dsurd}[2]{\frac{\partial #1}{\partial #2}}
\newcommand{\sdsurd}[2]{\partial #1/\partial #2}
\newcommand{\ddesurd}[2]{\frac{\partial^2 #1}{{\partial #2}^2}}
\newcommand{\vnabla}{\boldsymbol{\nabla}}
\newcommand{\vv}[1]{\mathbf{#1}}
\newcommand{\undemi}{\frac{1}{2}}
\newcommand{\cc}{\text{c.c.}}
\newcommand{\diverg}{\vnabla .}
\newcommand{\rhs}{right-hand-side }
\newcommand{\stensad}{S}
\newcommand{\paet}{P^*}
\newcommand{\phitherm}{\Phi_g}
\newcommand{\phar}{p_1}
\newcommand{\posc}{p_\text{osc}}
\newcommand{\pmoy}{p_\text{m}}
\newcommand{\conj}[1]{\overline{#1}}
\newcommand{\intp}{w}
\newcommand{\olchange}[1]{#1}

\begin{document}
%
\begin{frontmatter}
  \title{A simple model of ultrasound propagation in a cavitating
    liquid. Part I: Theory, nonlinear attenuation and traveling wave
    generation.}
  
  
  \author[EMAC]{O. Louisnard\corref{CORRESPONDANT}}
  \ead{louisnar@enstimac.fr}
  \cortext[CORRESPONDANT]{Corresponding author}
  \address[EMAC]%
  {Centre RAPSODEE, FRE CNRS 3213, Universit\'e de Toulouse, Ecole des Mines d'Albi, %
    \\  81013 Albi Cedex 09, France}
  

  \begin{abstract}
    The bubbles involved in sonochemistry and other applications of
    cavitation oscillate inertially. A correct estimation of the wave
    attenuation in such bubbly media requires a realistic estimation
    of the power dissipated by the oscillation of each bubble, by
    thermal diffusion in the gas and viscous friction in the liquid.
    Both quantities and calculated numerically for a single inertial
    bubble driven at 20~kHz, and are found to be several orders of
    magnitude larger than the linear prediction. Viscous dissipation
    is found to be the predominant cause of energy loss for bubbles
    small enough. Then, the classical nonlinear Caflish equations
    describing the propagation of acoustic waves in a bubbly liquid
    are recast and simplified conveniently. The main harmonic part of
    the sound field is found to fulfill a nonlinear Helmholtz
    equation, where the imaginary part of the squared wave number is
    directly correlated with the energy lost by a single bubble. For
    low acoustic driving, linear theory is recovered, but for larger
    drivings, namely above the Blake threshold, the attenuation
    coefficient is found to be more than 3 orders of magnitude larger
    then the linear prediction. A huge attenuation of the wave is thus
    expected in regions where inertial bubbles are present, which is
    confirmed by numerical simulations of the nonlinear Helmholtz
    equation in a 1D standing wave configuration.  The expected strong
    attenuation is not only observed but furthermore, the examination
    of the phase between the pressure field and its gradient clearly
    demonstrates that a traveling wave appears in the medium.
  \end{abstract}
  
\end{frontmatter}

\section{Introduction}

The complexity and large variety of spatial and temporal scales
involved in acoustic cavitation make difficult the derivation of a
full theoretical model, accounting for the coupled effects between the
bubble field and the sound field. Nevertheless, considerable progress
has been made in the last decade. Theoretical studies in the context
of single bubble sonoluminescence have allowed to restrict the ambient
size of the bubbles in the micron range, owing to surface
instabilities
\cite{hilgenlohsebrenner96,haoprosperetti99,yuan2001,storey2001,%
  brenner2002,linstoreyszeri2002,mettin2007}. This has been confirmed
by 20 kHz experiments, both in single bubble \cite{holtgaitan96} and
multi-bubble configurations (see Ref.~\cite{mettin2005} and references
therein).

Besides, a large collection of experimental observations have revealed
that radially oscillating bubbles in high-intensity acoustic fields
tend to self-organize into bubble structures, which shapes depend on
the experimental configuration, with possibly two structures or more
appearing simultaneously in different zones of the liquid
\cite{mettinappel2002,mettinkochkrefting2002,moussatovgrangerdubus2003,%
  appelkoch2004,mettin2005}. The shape of such structures is strongly
correlated with the fundamental issue of the translational motion of
the bubbles under Bjerknes forces, which have been reconsidered in the
context of strongly nonlinear inertial radial oscillations and
traveling waves \cite{akhatov97une,akhatov97deux,mettin2007}.

On this basis, the action of the acoustic field on the organization of
inertial bubbles has been satisfactorily described in various
configurations by particle models
\cite{parlitz99,mettinappel2002,appelkoch2004,mettin2007}, by
calculating the forces exerted on the bubbles directly from their
nonlinear dynamics.  Assuming a simple shape of the sound field, some
bubble structures have been remarkably caught by this method.
However, the correct prediction of other structures was found to be
more difficult, mainly because, as suggested by Mettin
\cite{mettin2005}, the local sound field might have a complicated
shape, which cannot be inferred without describing correctly the
acoustic field in the medium.

The backward effect of inertial bubbles on the propagation of acoustic
waves remains mainly unexplored.  \olchange{The main physical effects
  of the bubble radial oscillations on sound waves can be easily
  understood qualitatively. Bubbles are mechanical oscillators so that
  wave dispersion is expected. They oscillate non linearly for large
  amplitude drivings, so that waves should be nonlinear. Finally, they
  dissipate mechanical energy by various processes, which should
  produce wave attenuation. The problem has been attacked in the early
  work of Foldy \cite{foldy} who considered linear scattering of waves
  by an arbitrary statistical distribution of scatterers, and obtained
  a linear dispersion relation. The application of this theory to the
  specific case of linear sound waves in bubbly liquids has been
  considered in Refs.~\cite{cartensenfoldy,morsefesh}. A key feature
  in Foldy's approach is that for a sufficiently dilute bubbly
  mixture, each bubble behaves as if it were excited by the
  statistical average pressure field, which allows to cast aside the
  difficult issue of bubbles pairwise interaction. An intuitive
  justification of this approach can be found in
  Refs.~\cite{commprosper,prosperiutam}. 

  The assumption of small amplitude waves has been relaxed by
  Iordansky~ \cite{iordansky} and simultaneously by
  van~Wijngaarden~\cite{vanwijn65,vanwijn68} by a semi-empirical
  volume-averaging of the bubbly liquid equations, which is closed by
  a Rayleigh equation, where, as suggested by Foldy's work, the
  driving pressure term is the local average pressure field. The model
  obtained has allowed the study of nonlinear dispersive waves. The
  latter are classically described by the Korteweg-de Vries equation
  \cite{whithambook}, and the reduction of van~Wijngaarden model to
  the latter for moderate amplitudes has been studied by various
  authors both theoretically
  \cite{vanwijn68,vanwijn72,kuznetsov76,prosperetti91} and
  experimentally \cite{noordzij,kuznetsov78,vanwijn95}}.

The popular Caflish model~\cite{caflisha} \olchange{is a rigorous
  generalization of Foldy's theory to the nonlinear case and yields a
  simplified version of van Wijngaarden model, as far as the bubbly
  liquid is dilute enough. The latter hypothesis has the important
  corollary that the mean velocity of the mixture is infinitely small,
  so that the momentum conservation equation coincide with the one of
  linear acoustics [see Eq.~(\ref{cafqdm}) below]. A physical
  discussion of the latter feature can be found in
  Refs.~\cite{commprosper,prosperiutam}.}

Under the linear approximation, the Caflish model reduces to the
famous dispersion relation of Foldy, which can be extended to
calculate a wave attenuation coefficient, accounting for dissipation
by a linearly oscillating bubble \cite{commprosper} \olchange{and to
  polydisperse bubbles size distributions. Linearization allows a
simple description of the sound field by an Helmholtz equation, and
has been used in studies of the coupling between wave propagation and
the bubble field. The gain obtained by simplifying the wave equation
allows a complex description of its coupling with the bubble
population evolution, spatially and along the size axis.  Following
such an approach, Kobelev \& Ostrovski \cite{kobelevostro89} have
proposed an elegant model of self-action of low amplitude sound waves
in bubbly liquids, accounting for the bubble drift under the action of
primary Bjerknes forces and bubble coalescence favored by secondary
Bjerknes forces. Although the wave equation in this study was linear,
the global model was nonlinear, owing to the dependence of the wave
number on the varying bubble density, which conversely evolves
non linearly with the sound field. Specific solutions under different
hypothesis could catch the experimentally observed self-transparency, self-focusing of sound waves in bubbly liquids, and destabilization
of homogeneous bubble distributions. The latter instability has also
been demonstrated in Ref.~\cite{akhatov94} by a similar approach, but
involving a slightly different physics.}

\olchange{The attenuation of sound waves by oscillating bubbles
  remains normally weak for linear waves, except when the bubbles are
  close to the resonant size \cite{foldy,silberman,commprosper}, which
  is the main cause of sound extinction considered in
  Ref.~\cite{kobelevostro89}. Since low-frequency inertial cavitation
  involves bubbles much smaller than the resonant size
  \cite{mettin2005}, the use of the linear theory of
  Ref.~\cite{commprosper} predicts an abnormally low attenuation,
  compared to experimental data \cite{camposdubus2005}. This is not
  astonishing} since inertial bubbles typically suffer a ten-fold
expansion of their radius and are expected to dissipate more energy
than predicted by linear theory. \olchange{Despite the latter
  restriction, the linear dispersion relation has often been used to
  predict attenuation of strong cavitation fields, because it allows
  the description of the problem by a linear Helmholtz equation, which
  is easy to solve,} and allows harmonic response simulations
\cite{dahnke99,dahnke99_II,servant2003,%
  mettinkochlauterbornkrefting2006}. \olchange{Moreover, the use of a
  complex wave number provided by the linear dispersion relation in an
  Helmholtz equation somewhat masks the fact that one physical origin
  of wave attenuation by the bubbles is the energy dissipated by the
  latter. The latter point has been nicely addressed by Rozenberg
  \cite{rozenbergchap71}, who restated the problem of attenuation of a
  traveling wave by a cavitation zone in terms of energy
  conservation, without resorting to the linear hypothesis. The latter
  study made use of an empirical expression, fitted on experimental
  results, between the power dissipated by cavitation bubbles and the
  wave intensity. Doing so, realistic attenuated intensity profiles
  near the emitter could be calculated simply, and experimentally
  observed self-attenuation of the wave could be accounted
  for.}

\olchange{The last remarks suggests that the relaxation of the linear
  hypothesis is necessary to correctly predict attenuation by inertial
  bubbles, so that one should revert to the original fully nonlinear
  form of the Caflish model. However,} although valid \olchange{for
  any wave amplitude, the latter} remains intractable for large
multi-dimensional geometries, since it requires time-dependent
simulations, and presents convergence problems in the range of
inertial cavitation, \olchange{even in 1D
  \cite{louisnardthese,vanhille2009}}.  Thus, an intermediate model,
simple enough to be numerically tractable, but properly accounting for
the true energy dissipation by inertial bubbles, is necessary.

\olchange{The motivation of this work is the derivation of such a
  reduced model, and can be viewed as a systematic formalization of
  Rozenberg's approach \cite{rozenbergchap71}, based on the nonlinear
  Caflish model. The present paper extends the ideas formerly
presented in Ref.~\cite{louisnardICU2009} } and is organized as
follows. In section 2, we recast the fully nonlinear Caflish equations
into a mechanical energy balance equation, where we express explicitly
the energy lost by the bubbly liquid on average over an oscillation
period, as functions of period-averaged quantities of a single bubble
dynamics.  This energy loss is then computed numerically, by
simulating a bubble radial dynamics equation over a typical parameter
range, including the range of inertial cavitation involved in
cavitation and sonochemistry experiments. In section~3, we then seek a
reduction of the Caflish equations for the main harmonic component of
the acoustic field, involving the energy dissipation calculated in
section~2. Finally, in section~\ref{secresults}, the resulting
nonlinear Helmholtz equation is solved numerically in a 1D
configuration, and a detailed analysis of the obtained wave profiles
is performed. The implications of the present results on the primary
Bjerknes forces and 2D simulations of classical experimental
configurations are deferred in a companion paper.

\section{Theory}
\subsection{Caflish equations}
The Caflish model \cite{caflisha} describes the propagation of an
acoustic wave of arbitrary amplitude in a bubbly liquid described as a
continuum, which means that the radial oscillations of all the bubbles
pertaining to an elementary small volume of mixture located at a
spatial point $\vr$ can be described by a continuous spatio-temporal
radius function $R(\vr,t)$. \olchange{The first two equations of the model
correspond to mass and momentum conservation in the mixture:}
\begin{eqnarray}
  \label{cafmass}
  \frac{1}{\rho_l c_l^2}\dsurd{p}{t} + \vnabla.\vv{v} &=&
  \displaystyle\dsurd{\beta}{t}, \\
  \label{cafqdm}
  \rho_l\dsurd{\vv{v}}{t} + \vnabla p &=& 0.
\end{eqnarray}
In the above equations, $p(\vr,t)$ is the acoustic pressure field,
$\vv{v}(\vr,t)$ the velocity field, $\rho_l$ the liquid density, $c_l$
the sound speed in the liquid, and $\beta(\vr,t)$ is the instantaneous
void-fraction, which, assuming a mono-disperse distribution of the
bubbles, can be defined by:
\begin{equation}
  \label{defbeta}
  \beta(\vr,t) = N(\vr) \frac{4}{3}\pi R(\vr,t)^3,
\end{equation}
where $N(\vr)$ is the local bubble density. The latter is assumed
time-independent, or at least almost constant on the time scale of the
oscillations.  Despite the set of equations (\ref{cafmass})
and~(\ref{cafqdm}) is very similar to the equations of linear
acoustics, the presence of the right-hand-side term of
Eq.~(\ref{cafmass}) renders the whole model nonlinear.  Following the
procedure classically used for linear acoustics, these two equations
can be easily recast into an equation of energy conservation, by
multiplying (\ref{cafmass}) by $p$ and~(\ref{cafqdm}) by $\vv{v}$:
\begin{equation}
  \label{eqacoustenerliq}
  \dsurd{}{t}\left(\undemi\frac{p^2}{\rho_l c_l^2}+\frac{1}{2}\rho_l v^2
  \right) + \diverg\left(p\vv{v} \right) = N p\dsurd{V}{t},  
\end{equation}
where $V(\vr,t)$ denotes the instantaneous volume of the bubbles
located at $\vr$. The time derivative in the left-hand-side (LHS) of
this equation represent the time-variations of the acoustic energy
density, which is the sum of kinetic energy and potential
compressional energy of the pure liquid. The second LHS-term is the
divergence of the acoustic intensity $p\vv{v}$. The \rhs (RHS), which
would be zero for a linear wave propagating in the pure liquid,
represents the mechanical power exchanged between the acoustic wave
and the bubbles. As will be seen below, part of this energy is
irreversibly dissipated along the radial oscillations of the bubbles,
which is the physical origin of the acoustic wave attenuation.

\subsection{Bubble dynamics}
The bubble radial motion equation can be described by a radial
dynamics equation. The Caflish model in its original form uses a
inviscid Rayleigh-Plesset equation with isothermal behavior of the
bubble, in which the infinite driving pressure field is the mean local
acoustic pressure field $p(\vr,t)$. In the present study, we want to
examine the energy dissipation by heat transfer between the bubble
interior and the liquid, and by viscous friction in the radial motion
of the liquid around the bubble. We therefore leave the bubble
pressure~$p_g$ unspecified for now, and add the classical viscous term
in the Rayleigh-Plesset equation. Besides, since surface tension plays
a preponderant role in inertial cavitation
\cite{akhatovgumerov97,hilgenbrennergrosslohse98,louisnardgomez2003,%
  louisnard2008}, we also added the correction accounting for the
latter effect, so that the bubble dynamics is given by:
\begin{equation}
  \label{rayleigh}
  \rho_l\left(R\ddot{R} + \frac{3}{2}\dot{R}^2 \right) = 
  p_g-\frac{2\sigma}{R}-4\mu_l\frac{\dot{R}}{R} - p,
\end{equation}
where $\mu_l$ is the liquid dynamic viscosity, and $\sigma$ the
surface tension.  All the quantities $R$, $p_g$ and $p$ in this
equation are spatio-temporal fields, depending on both $\vr$ and $t$,
so that the time derivatives represented by over-dots in this equation
must be understood as partial derivatives $\sdsurd{}{t}$ at $\vr$
constant. \olchange{We did not add any corrections accounting for
  liquid compressibility, in order to keep a reasonably simple model.
  We defer the discussion of this choice to the conclusion section.}

For further use in the paper, we recall that when a bubble is driven
by a sinusoidal pressure field $p = p_0\left[1-\paet\sin (2\pi f t
  )\right]$ around the ambient pressure $p_0$, its oscillations become
inertia-controlled and involve a strong collapse when the driving
pressure amplitude is above the Blake threshold
\cite{akhatovgumerov97,hilgenbrennergrosslohse98,%
  louisnardgomez2003}:

\begin{equation}
  \label{blake}
  \paet_B = 1+ \left(\frac{4}{27}
    \frac{\stensad^3}{1+\stensad} \right)^{1/2},
\end{equation}
where $S=2\sigma /(p_0 R_0)$ is the dimensionless Laplace tension and
$R_0$ the bubble ambient radius. Such an oscillation regime,
historically termed as ``transient cavitation'', is now classically
named as ``inertial cavitation'' \cite{matula99,brenner2002}.

\subsection{Energy dissipation per bubble}
\label{secpis}
In order to get an energetic interpretation of the bubble radial
motion, equation (\ref{rayleigh}) can be multiplied by the time
derivative of the bubble volume $\sdsurd{V}{t}$, and noting that
$\rho_l(R\ddot{R} + \frac{3}{2}\dot{R}^2)\times\sdsurd{V}{t}$ is the
time-derivative of the radial kinetic energy of the liquid~$K_l =
2\pi\rho_l R^3\dot{R}^2$, we obtain:
\begin{equation}
  \label{eqKbubble}
     \dsurd{}{t}\left(K_l + 4\pi R^2\sigma \right) 
 =  - 16\pi\mu_l R\dot{R}^2
   -p\dsurd{V}{t} 
 + p_g\dsurd{V}{t}.
\end{equation}
This equation is strictly equivalent to the Rayleigh equation, and is
the expression of the theorem of kinetic energy applied to the liquid
surrounding the bubble. The parenthesis in the LHS of
(\ref{eqKbubble}) represents the sum of the kinetic energy of the
radially moving liquid and the interfacial potential energy.

The first term in the RHS of Eq.~(\ref{eqKbubble}) is the power
irreversibly lost by internal viscous friction within the liquid as it
moves radially. 

The second term in the RHS is the power transferred from the acoustic
field to the liquid surrounding the bubble, and can be viewed as the
energy source available to drive the bubble oscillations and the
radial motion of the liquid around. When multiplied by the number of
bubbles per unit-volume, this term is similar to the \rhs of
Eq.~(\ref{eqacoustenerliq}) with the opposite sign, which clearly
indicates how energy is transferred between the driving acoustic field
and the radially oscillating bubble.

Finally, the last term in the RHS of (\ref{eqKbubble}) is the
mechanical power done by the gas on the liquid, and could be expressed
as the time-derivative of a compressional energy $-\sdsurd{E_p}{t}$ in
the case of a barotropic relation between the bubble pressure $p_g$
and volume $V$ (for example assuming an isothermal \cite{caflisha} or
adiabatic evolution of the gas).  However, in the general case where
heat flows irreversibly between the bubble interior and the liquid,
this term cannot be expressed as the time-derivative of a potential
function, and we now detail how this term is linked to dissipation of
energy over a whole oscillation cycle of the bubble.

In what follows, we will assume periodic oscillations of all the
fields. Averaging Eq.~(\ref{eqKbubble}) over one cycle, the
time-derivative in the left side cancels and we get:
\begin{equation}
  \label{eqKbubblemoy}
  \left<- p\dsurd{V}{t} \right> = \pith + \piv,  
\end{equation}
where the two bubble dynamics-dependent average quantities $\pith$ and
$\piv$ read:
\begin{eqnarray}
  \label{defpith}
  \pith &=&  \frac{1}{T} \int_0^T - p_g \dsurd{V}{t}\; d t , \\
  \label{defpiv}
  \piv &=&  \frac{1}{T} \int_0^T 16\pi\mu_l R\dot{R}^2 \;d t.
\end{eqnarray}
The quantity $\piv$ defined by (\ref{defpiv}) is clearly positive, and
is the period-averaged power loss by viscous friction in the liquid.

A clear interpretation of $\pith$ can be obtained by applying the
first principle of thermodynamics to the whole bubble content, which
yields integral (\ref{defpith}) as:
\begin{equation}
  \label{premierprincipe}
  \pith = \frac{1}{T}\int_0^T \frac{d(U_g+K_g)}{d t}\;d t 
  - \frac{1}{T}\int_0^T \dot{Q}\;d t,
\end{equation}
where $U_g$ and $K_g$ depict the internal energy and kinetic energy,
respectively, of the whole gas in the bubble, and $\dot{Q}$ is the
heat gained by the bubble over one cycle.  The first integral in the
\rhs of (\ref{premierprincipe}) is zero for a periodic motion, so that
$\pith=-\left<\dot{Q}\right>$ is just the net heat lost by the bubble over
one oscillation cycle.

Equation (\ref{eqKbubblemoy}) has therefore the following physical
meaning: the energy transferred by the acoustic field to the bubble
over one acoustic period is dissipated by two processes: the heat flow
from the bubble toward the liquid and the viscous friction in the
liquid radial motion.


The integrals $\piv$ and $\pith$ can be evaluated numerically by
solving the bubble dynamics equation (\ref{rayleigh}) for an arbitrary
single bubble of ambient radius $R_0$ excited by a sinusoidal forcing
$p = p_0\left[1-\paet\sin (2\pi f t )\right]$, possibly varying 
the acoustic parameters $\paet$ and~$f$, the bubble ambient radius $R_0$,
and the properties of the liquid and the gas.  In this paper we will
restrict to air bubbles in water at ambient pressure excited at 20~kHz
and take: $p_0=101300$ Pa, $\rho_l=$~1000 kg/m$^3$, $\mu_l=$~10$^{-3}$
Pa.s, $\sigma =$~0.0725 N.m$^{-1}$.  The bubble ambient radius $R_0$
and driving pressure amplitude $P$ will be varied within a range of
interest. More results involving, among others, the effect of the
frequency and the type of gas will be given elsewhere
\cite{olprepare}.

Since $\pith$ represents the net heat flow leaving the bubble, thermal
diffusion in the bubble interior must be properly accounted for in our
simulations, at least in an approximate manner. To that aim, the bubble
interior is modeled by an approximate energy conservation equation
based on a thermal diffusion layer
\cite{toegel2000,storeyszeri2001,matulahilmo2002}.  Water evaporation
and condensation at the bubble interface is also taken into account by
a similar method, as described in the same references.  The latter
refinement may be important since the presence of water vapor in the
bubble is known to decrease the temperature collapse \cite{toegel2000}
and therefore influences the estimation of $\dot{Q}$. To solve the
bubble radial dynamics, the variables are non-dimensionalized by:
\begin{equation*}
  t^* = \omega t, \qquad R^* = \frac{R}{R_0},
 \qquad p_g^* = \frac{p_g}{p_0}, 
 \qquad \Pi_{\text{v,th}}^* = \frac{\Pi_{\text{v,th}}}{p_0V_0\omega},
\end{equation*}
and the dimensionless dissipation functions $\pithet$ and $\pivet$ are
calculated numerically by:
\begin{eqnarray}
  \label{defpithnum}
  \pithet &=& 
  \frac{1}{2\pi}\left(1+\frac{2\sigma}{p_0R_0} \right)
  \int_0^{2\pi} p_g^* \frac{d V^*}{d t^*} d t^*, \\
  \label{defpivnum}
  \pivet  &=& \frac{6}{\pi}\frac{\omega\mu_l}{p_0}
  \int_0^{2\pi} R^* \left(\frac{d R^*}{d t^*} \right)^2 d t^*.
\end{eqnarray}
For comparison purposes, we recall that assuming linear oscillations
of the bubble, the equation of radial dynamics can be linearized by
setting
\begin{equation}
  \label{defrlin}
   R^*(t) = 1+\frac{1}{2}\left(Xe^{it^*} + \cc \right),
\end{equation}
where the complex amplitude can be obtained analytically,
accounting rigorously for thermal effects \cite{devin,
  prosperetti77th, prosperetti91}. Introducing (\ref{defrlin}) in
(\ref{defpithlin})-(\ref{defpivlin}), $\pithet$ and $\pivet$ can be
obtained analytically, and we obtain:
\begin{eqnarray}
  \label{defpithlin}
  \pithetlin &=& \frac{3}{2} \left(1+\frac{2\sigma}{p_0R_0}
  \right)\Im(\phitherm) |X|^2, \\
  \label{defpivlin}
  \pivetlin &=& \frac{6\mu_l\omega}{p_0} |X|^2,
\end{eqnarray}
where $\phitherm$ is a complex dimensionless number which can be
expressed in terms of the gas thermal P\'eclet number
$\text{Pe}_\text{th}=R_0^2\omega/\chi_g$, where $\chi_g$ is the
thermal diffusivity of the gas in ambient conditions
\cite{prosperetti77th,prospercrumcomm}.

\begin{figure}[ht]
  \centering
  \includegraphics[width=\linewidth]{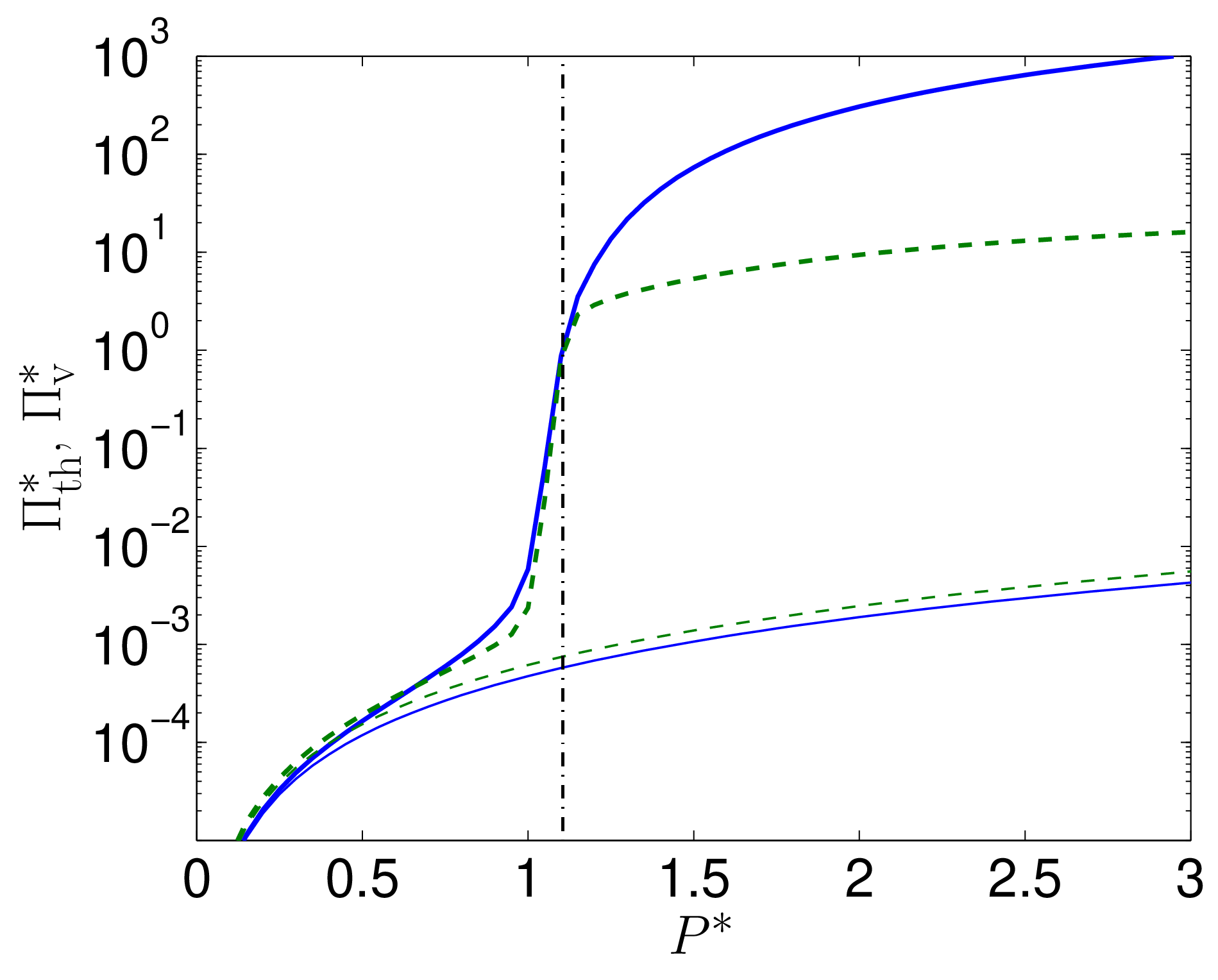}
  \caption{ Dimensionless power dissipated by an argon bubble of
    ambient radius $R_0 = 3$ $\mu$m in water, at 20 kHz: by viscosity
    $\pivet$ [thick solid line, from Eq.~(\ref{defpivnum})]; by
    thermal diffusion, $\pithet$ [thick dashed line, from
    Eq~(\ref{defpithnum})].  The thin lines are the corresponding
    values obtained from linear theory,
    Eqs.~(\ref{defpithlin})-(\ref{defpivlin}) (solid: $\pivetlin$;
    dashed: $\pithetlin$). The vertical
    dash-dotted line represents the Blake threshold calculated by
    Eq.~(\ref{blake}) .}
 \label{figpidissr3}
\end{figure}

Figure~\ref{figpidissr3} displays the values calculated for
$\pivet$ and $\pithet$ for an air bubble of ambient radius $R_0
= 3$ $\mu$m driven at 20~kHz in water at ambient pressure and
temperature. First, it is seen that the power dissipated either by
viscous friction (thick solid line) or by thermal diffusion (thick
dashed line) quickly rises in the neighborhood of the Blake threshold
(where approximately $\pivet\simeq\pithet\simeq 1$), well above their
value predicted by linear theory (between 5 and 6 orders of
magnitude). This clearly demonstrates the need for exact nonlinear
bubble dynamics to calculate realistic values of the energy dissipated
by inertial bubbles.

Another interesting feature is that, for the parameters used in
Fig.~\ref{figpidissr3}, viscous dissipation becomes much larger than
the thermal one (more than one order of magnitude), for driving
pressures above the Blake threshold, whereas linear theory predicts
the opposite in this parameter range. Viscous dissipation in the
liquid is thus found to largely predominate over the thermal one for 3
$\mu$m inertial bubbles. 

\olchange{It is also interesting to interpret these results in the
  light of the experimental data reported by Rozenberg
  \cite{rozenbergchap71}, who fitted the volumic power dissipated in
  the cavitation zone by the following function of sound intensity
  $I$:
  \begin{equation}
    {\mathcal P} = \left\{
      \begin{array}{ll}
        A (I-I_t)^2, & I > I_t \\
        0, & I \le I_t 
      \end{array}
    \right.,
  \end{equation}
  where $I_t$ is the intensity cavitation threshold. Identifying
  ${\mathcal P}$ with $N(\pith+\piv)$, noting that sound intensity $I$
  scales as $P^2$ for traveling waves, and identifying the cavitation
  threshold with Blake threshold, Rozenberg's result suggests that
  $\pith + \piv$ would scale as $P^2-P_B^2$ for $P> P_B$, and would be
  0 under the threshold. This is almost consistent with our results,
  except that redrawing Fig.~\ref{figpidissr3} with linear scale (not
  shown) would reveal a linear dependence rather than a quadratic one.
  However, on one hand, Rozenberg's results apply to 500~kHz fields,
  and on the other hand, it is highly probable that the bubble density
  $N$ also depends on the local sound field, despite we will consider
  $N$ constant above the Blake threshold in the model developed below
  (see Sec.~\ref{secresults}).}

\begin{figure}[ht]
  \centering
  \includegraphics[width=\linewidth]{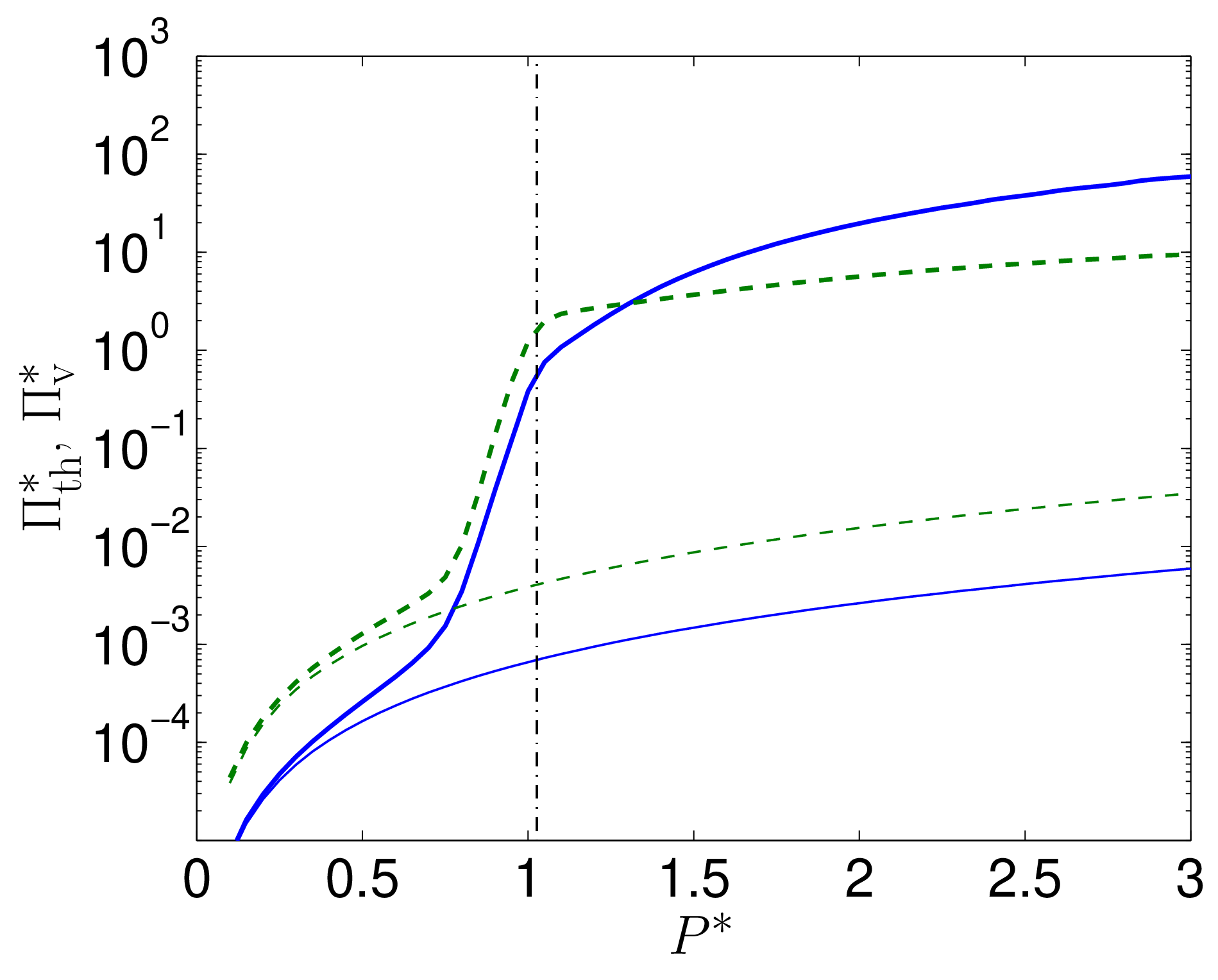}
  \caption{Same as Fig.~\ref{figpidissr3} for a 8~$\mu$m bubble.}
 \label{figpidissr8}  
\end{figure}

We repeated the same calculation for a 8~$\mu$m bubble
(Fig.~\ref{figpidissr8}). The scale is chosen identical as
Fig.~\ref{figpidissr3} in order to make the comparison easier. The
same conclusions apply except that the increase of viscous dissipation
$\pivet$ over the Blake threshold is lower than for the 3 $\mu$m
bubble, and remains of the same order of magnitude as $\pith$ for
moderate driving pressures.

\begin{figure}[ht]
  \centering
  \includegraphics[width=\linewidth]{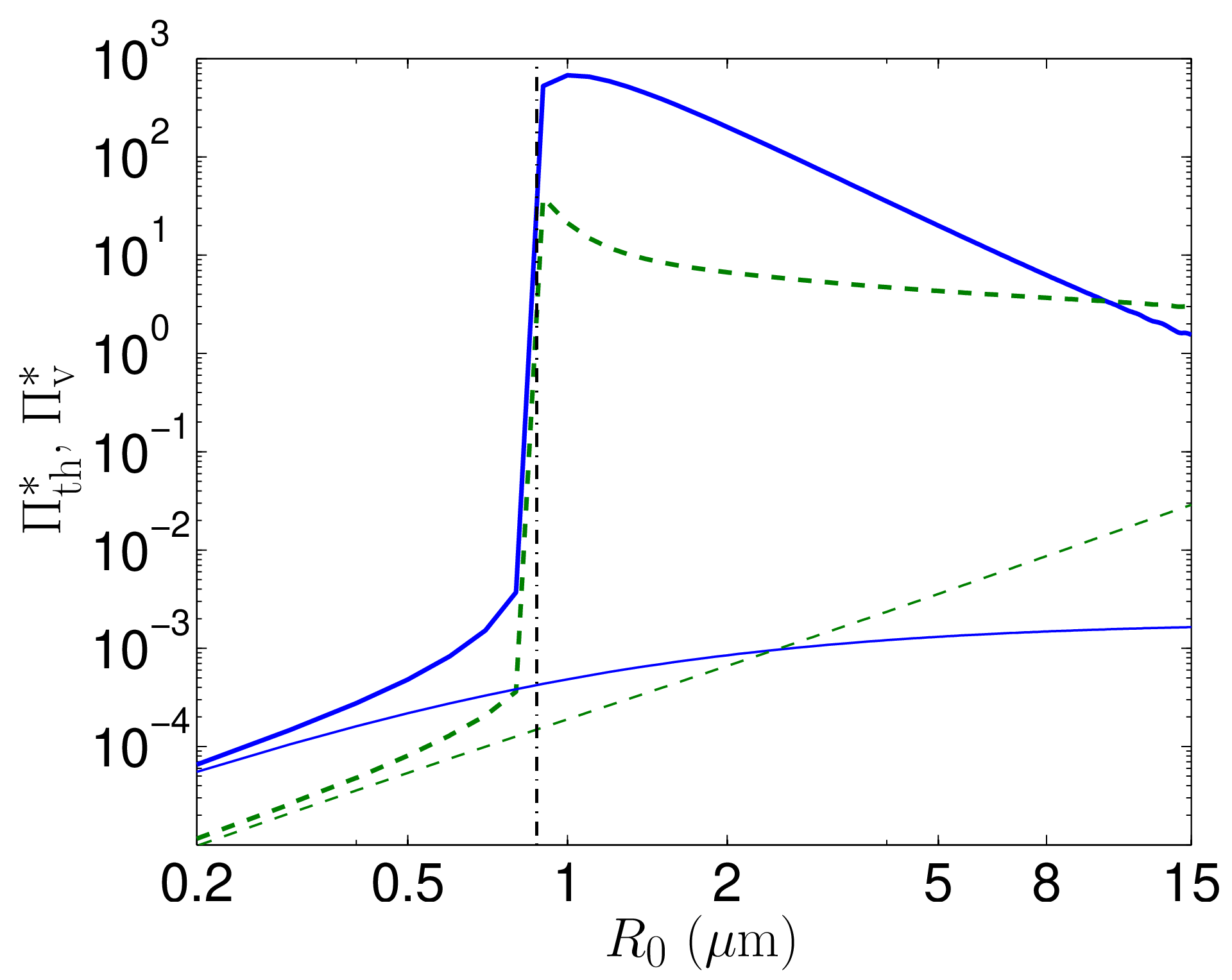}
  \caption{Same as Figs.~\ref{figpidissr3} and~\ref{figpidissr8}, but
    varying $R_0$ for $\paet=1.5$. The vertical dash-dotted line
    represents the Blake threshold.}
 \label{figpidissp1_5}
\end{figure}

To assess more clearly the dependence of $\piv$ and $\pith$ on the
ambient radius $R_0$, we calculated $\piv$ and $\pith$ at constant
$\paet= 1.5$, but varying $R_0$.  The result is displayed in
Fig.~\ref{figpidissp1_5}. Viscous dissipation $\piv$ is much larger
than thermal dissipation $\pith$ just above the Blake threshold, and
decreases below $\pith$ only above $R_0 \simeq 10$~$\mu$m.

More curves like the ones of Figs.~\ref{figpidissr3},
\ref{figpidissr8} and \ref{figpidissp1_5} could be drawn, but we can
summarize the comparison of $\piv$ and $\pith$ above the Blake
threshold as follows: $\piv$ predominates for larger drivings and
smaller bubbles, while the opposite is true for larger bubbles and
smaller drivings. Since for large drivings, surface instabilities
maintain the ambient radii of inertial bubbles in a small interval
just above the Blake threshold
\cite{hilgenlohsebrenner96,gaitanholt99,yuan2001}, this suggests that
viscous friction would be the predominant dissipation phenomenon in
cavitation clouds.

The real power dissipated by an inertial bubble is therefore larger
than the one predicted by linear theory by several orders of
magnitude. We therefore expect the real wave attenuation in a liquid
containing inertial bubbles (above the Blake threshold) to be much
higher than the value calculated by linear theory. We will quantify
this point in Sec.~\ref{secmodel}.

Although the above results are sufficient to carry on the development
of our model, it is instructive to close this section by relating the
dissipation functions $\pith$ and $\piv$ to the conservation of
acoustic energy, generalizing the conservation equation proposed in
the original paper of Caflish et al.  \cite{caflisha}.

\subsection{Conservation of  energy in the bubbly liquid }
The term $p\sdsurd{V}{t}$ can be eliminated between
equations (\ref{eqacoustenerliq}) and (\ref{eqKbubble}), by
multiplicating the latter by $N$, to obtain a global energy
conservation equation of the bubbly liquid:
\begin{equation}
  \label{energyfinal}
  \begin{split}
    & \dsurd{}{t}\left(
      \undemi \frac{p^2}{\rho_l c_l^2}+\frac{1}{2}\rho_l v^2 
      + N K_l +4\pi N\sigma R^2 \right)\\
    & \quad + \diverg\left(p\vv{v} \right) = 
    N p_g\dsurd{V}{t} - N 16\pi\mu_l R\dot{R}^2.
      \end{split}
\end{equation}
Equation (\ref{energyfinal}) represents the conservation of mechanical
energy of the bubbly liquid:

\begin{itemize}
\item $p^2/(2\rho_l c_l^2)$ is the elastic potential energy stored by the
  pure liquid involved in the propagation of the wave,
\item $\rho_l v^2/2$ is the kinetic energy per unit volume of the
  pure liquid involved in the propagation of the wave,
\item $N K_l$ is the kinetic energy per unit volume of the
  liquid in its radial motion around the bubbles,
\item $4\pi N \sigma R^2$ is the interfacial potential energy per unit
  volume,
\item $p\vv{v}$ is the acoustic intensity, or flux density of
  mechanical energy. It is supplied at a vibrating boundary in contact
  with the bubbly liquid, typically by the oscillating motion of the
  sonotrode \cite{louisnardgonzalez2009}.
\end{itemize}

In what follows, we will assume periodic oscillations of all the
fields. Averaging Eq.~(\ref{energyfinal}) over one cycle, the
time-derivative in the left side cancels and we get:
\begin{equation}
  \label{energyfinalaverage}
    \diverg\left<p\vv{v} \right> = -N \left(\pith+\piv \right).
\end{equation}
Equation~(\ref{energyfinalaverage}) is the conservation of mechanical
energy averaged over one period of oscillation, and has a clear
physical interpretation: the balance between the acoustic energy
leaving a volume of bubbly liquid and the one reaching it is always
negative, owing to thermal loss in the bubble and viscous friction in
the radially moving liquid.  Each bubble therefore appears as a
dissipator of acoustic energy, owing to these two phenomena.  The
physical origin of wave attenuation is thus self-contained in the
Caflish model, even for nonlinear oscillations, provided that a
correct model is used to describe thermal diffusion in the bubble
interior.  Caflish and co-workers proposed a conservation equation
similar to (\ref{energyfinal}), disregarding viscosity and assuming
isothermal oscillations, in which case mechanical energy is conserved
\cite{caflisha}. \olchange{It should also be noted that
  Eq.~(\ref{energyfinalaverage}) reverts exactly the equation solved
  in 1D by Rozenberg \cite{rozenbergchap71} in the case of purely
  traveling waves, but in the latter work, the dissipated power was
  fitted from experimental data, rather than being calculated ab
  initio from single bubble dynamics as done in the present work. }

\section{The model}
\label{secmodel}
\subsection{Intuitive approach}
We first recall that the velocity field can be eliminated
between Eq.~(\ref{cafmass}) and (\ref{cafqdm}) to yield an equation
involving only the pressure field \cite{caflisha,commprosper}:
\begin{equation}
  \label{cafprop}
  \nabla^2 p = \frac{1}{c_l^2}\ddesurd{p}{t} -\rho_l\ddesurd{\beta}{t}.
\end{equation}
Setting the pressure field $p$ as a mono-harmonic wave:
\begin{equation*}
  p(\vr,t) = \undemi \biggl( 
  P(\vr)e^{i\omega t} + \conj{P}(\vr)e^{-i\omega t} 
  \biggr),
\end{equation*}
the linearization of the above equation and of the bubble dynamics
equation allows to show that the complex field  $P$ fulfills an
Helmholtz equation:
\begin{equation*}
  \nabla^2 P + k^2 P = 0,
\end{equation*}
where the complex wave number is given by the linear dispersion
relation \cite{foldy,vanwijn65,caflisha,commprosper,prosperetti91}:
\begin{equation}
  \label{reldispers}
  k^2 = \frac{\omega^2}{c_l^2} + \frac{4\pi R_0 \omega^2 N}
  {\omega_0^2-\omega^2 +2i b\omega}.
\end{equation}
In Eq.~(\ref{reldispers}), $\omega_0$ is the resonance frequency and
$b$ the damping factor,  respectively given by
\begin{eqnarray}
  \label{defomega0}
  \omega_0^2 &=& \displaystyle\frac{p_0}{\rho R_0^2}
  \left[(1+\stensad)\Re(\phitherm)-\stensad \right], \\
  \label{defb}
  2b &=& \displaystyle
  \frac{p_0(1+\stensad)}{\rho\omega R_0^2}\Im(\phitherm) 
  + \frac{4\mu_l}{\rho R_0^2}.
\end{eqnarray}
It can be readily seen, that, even for sub-resonant bubbles ($\omega <
\omega_0)$, the wave number is complex because of the damping factor
$b$, which, as expected from the discussion in Sec.~\ref{secpis}, is
correlated with the heat loss from the bubble and the viscous friction
in the liquid. The imaginary part of the wave number represents the
attenuation factor of the wave, and can be easily calculated by
setting $k=k_r-i\alpha$ and identifying $k_r$ and $\alpha$ from
Eq.~(\ref{reldispers}). 

Generalizing this simple theory for inertial cavitation sounds
unrealistic, since the bubble dynamics cannot be reasonably linearized
for inertial oscillations. Thus all the fields are not mono-harmonic
anymore and the problem cannot be reduced to an Helmholtz equation.
However, for periodic oscillations, either linear or not, the
correlation between the energy dissipated by each bubble over one
cycle and the attenuation of the wave remains a universal principle,
formalized by Eq.~(\ref{eqKbubblemoy}), and constitutes the guideline of
the following derivation. 

We will therefore show that the first harmonic component of the field
(at the frequency $\omega$ of the driving) approximately follows an
Helmholtz equation, but whose wave number is directly expressed as
functions of the dissipation functions $\pith$ and $\piv$ presented in
the precedent section. This procedure allows to generalize the linear
model, in the sense that the time-variable is eliminated, but keeping
realistic values for the energy dissipated by inertial bubbles.

\subsection{Derivation of the model}
We decompose the pressure field into a sum of a time-average pressure
$\pmoy$, a first harmonic pressure $\phar$, oscillating at the
frequency of the ultrasonic source, and harmonic terms noted $\posc$,
that could be written as a Fourier series starting with a term at the
frequency $2\omega$: \footnote{We assume here for simplicity that
  there is no subharmonics or ultra-harmonic terms, but the following
  reasoning can always be generalized by taking time-averages over the
  largest period of the pressure field.}
\begin{equation}
  \label{pdecomp}
  p(\vr,t) = \pmoy(\vr) + \phar(\vr, t) + \posc(\vr,t).
\end{equation}
The first harmonic pressure field $\phar$ is expressed as:
\begin{equation}
  \label{defphar}
  \phar(\vr,t) = \undemi \biggl( 
  P(\vr)e^{i\omega t} + \conj{P}(\vr)e^{-i\omega t}
  \biggr),
\end{equation}
where over-lines denote complex conjugate. Next, we set $\intp$ the
primitive of the first harmonic pressure field:
\begin{equation}
  \label{defintp}
  \intp =  \undemi\frac{1}{i\omega} \biggl( 
  P(\vr)e^{i\omega t} - \conj{P}(\vr)e^{-i\omega t} 
  \biggr).
\end{equation}
Multiplicating the propagation equation (\ref{cafprop}) by
$\intp$ and averaging over one acoustic period yields:
\begin{equation}
  \label{cafpropmoy}
  \left<w\nabla^2 p \right>  =  
  \frac{1}{c_l^2}\left<w \ddesurd{p}{t} \right> -\rho_l\left<w\ddesurd{\beta}{t} \right>.
\end{equation}
Integrating by parts, using the definition of $\intp$, and the  fact
that all quantities are periodic, we obtain:
\begin{equation}
  \label{cafpropmoyintpart}
  \left<w\nabla^2 p \right>  =  
  -\frac{1}{c_l^2} \left<\phar \dsurd{p}{t} \right> 
  + \rho_l\left<\phar\dsurd{\beta}{t} \right>.  
\end{equation}
Using the decomposition (\ref{pdecomp}), it can be easily seen that
$\left<w\nabla^2 p \right>= \left<w\nabla^2 p_1 \right>$ and that
$\left<\phar\sdsurd{p}{t} \right>=0$. Besides, using
Eq.~(\ref{pdecomp}), the second term of the \rhs of
(\ref{cafpropmoyintpart}) can be expressed as:
\begin{equation*}
  \left<\phar\dsurd{\beta}{t} \right> = 
  \left<p\dsurd{\beta}{t} \right> 
  - \left<\posc\dsurd{\beta}{t} \right> ,
\end{equation*}
since $\left<\pmoy\sdsurd{\beta}{t} \right> =
\pmoy\left<\sdsurd{\beta}{t} \right> = 0$. We now make the empirical
assumption that $\left<\posc\sdsurd{\beta}{t} \right>$ is negligible.
%
%
A rigorous justification for this assumption is difficult in absence
of results on the respective orders of magnitude of $\phar$ and
$\posc$. However, unpublished measurements show that the latter is generally
one order of magnitude lower than the former, so that for now, we
assume that the assumption is justified. We therefore conclude that:
\begin{equation}
  \label{approxmono}
  \left<\phar\dsurd{\beta}{t} \right> \simeq
  \left<p\dsurd{\beta}{t} \right> 
\end{equation}
A physical interpretation of this approximate equation can be given by
looking at Eqs.~(\ref{eqacoustenerliq}), (\ref{eqKbubble}): it reverts
to consider that the interaction between the acoustic field and the
bubbles only occur through the first harmonic part of the field, and
that the bubble mainly responds to this first harmonic content. We
will term this hypothesis as ``first harmonic approximation'' (FHA).
From this assumption and the above derivation,
Eq.~(\ref{cafpropmoyintpart}) takes therefore the approximate form:
\begin{equation}
  \label{cafpropmoyapprox}
  \left<w\nabla^2 p_1 \right>  =  
   \rho_l\left<p\dsurd{\beta}{t} \right>,
\end{equation}
and using Eq.~(\ref{eqKbubblemoy}), Eq.~(\ref{cafpropmoyapprox}),
we finally obtain:
\begin{equation}
  \label{cafpropmoyapproxfin}
  \left<w\nabla^2 p_1 \right>  =  -N\left(\pith + \piv \right).
\end{equation}
We can now use the harmonic expressions (\ref{defphar}) and (\ref{defintp}) of
$\phar$ and $\intp$, to obtain:
\begin{equation*}
  \frac{i}{4\omega}\left(\conj{P}\nabla^2 P - P \nabla^2\conj{P}
  \right) = -\rho_l N \left(\pith+\piv \right),
\end{equation*}
and, dividing both sides of this equation by $|P|^2$, $P$ is finally found
to fulfill:
\begin{equation}
  \label{cafmoyfin}
  \Im\left(\frac{\nabla^2 P}{P} \right) = 
  2\rho_l\omega N \frac{\pith+\piv}{|P|^2},
\end{equation}
where $\Im$ denotes the imaginary part.  We therefore see that if $P$
were to fulfill an Helmholtz equation, the wave number would
necessarily satisfy following relation:
\begin{equation}
  \label{defk2i}
  \Im\left(k^2 \right) = 
  - 2\rho_l\omega N \frac{\pith+\piv}{|P|^2}.
\end{equation}
This equation is a generalization of the linear case represented by
Eq.~(\ref{reldispers}), but here, $\pith$ and $\piv$ can be estimated
from fully nonlinear bubble dynamics. By the way, it can be checked
after a few algebra that, linearizing $\pith$ and $\piv$,
Eq.~(\ref{defk2i}) yields the same results as taking the imaginary
part of the dispersion relation (\ref{reldispers}). For linear
oscillations, $\pith$ and $\piv$ scale as $|P|^2$ (see left part of
the curves in Figs.~\ref{figpidissr3} and \ref{figpidissr8}), so that
linear theory yields a value of $\Im\left( k^2 \right)$ independent of
the driving amplitude~$|P|$. This is no longer the case for nonlinear
oscillations and Eq.~(\ref{defk2i}) yields a value of $\Im\left(k^2
\right)$, which now depends on the local magnitude of the acoustic
pressure $|P|$.

The idea of the present paper is thus to use Eq.~(\ref{defk2i}) by
using the nonlinear values of $\pith$ and $\piv$ obtained in
Sec.~\ref{secpis} to calculate $\Im(k^2)$, and, relying on
Eq.~(\ref{cafmoyfin}), to introduce the latter in a nonlinear
Helmholtz equation:
\begin{equation}
  \label{helmholtzNL}
  \nabla^2 P + k^2\left(|P| \right) P = 0.
\end{equation}
Clearly, owing to the approximations made above, some additional terms
would appear in the exact equation fulfilled by $P$. However,
Eq.~(\ref{defk2i}) has the advantage to clearly link the attenuation
factor to the real dissipation of energy by the bubbles. Since it only
yields the imaginary part of~$k^2$, there remains the problem of
calculating its real part. For now, we still use the linear dispersion
relation to evaluate $\Re(k^2)$, and defer the discussion of this
approximation below:
\begin{equation}
  \label{defk2r}
  \Re(k^2) =  \frac{\omega^2}{c_l^2} + \frac{4\pi R_0 \omega^2 N}
  {\omega_0^2-\omega^2}.
\end{equation}
The attenuation coefficient and the real part of the wave number can now
be deduced from:
\begin{equation}
  \label{kglob}
  k=k_r-i\alpha,
\end{equation}
and by identification with (\ref{defk2i})-(\ref{defk2r}).

\begin{figure}[ht]
  \centering
  \includegraphics[width=\linewidth]{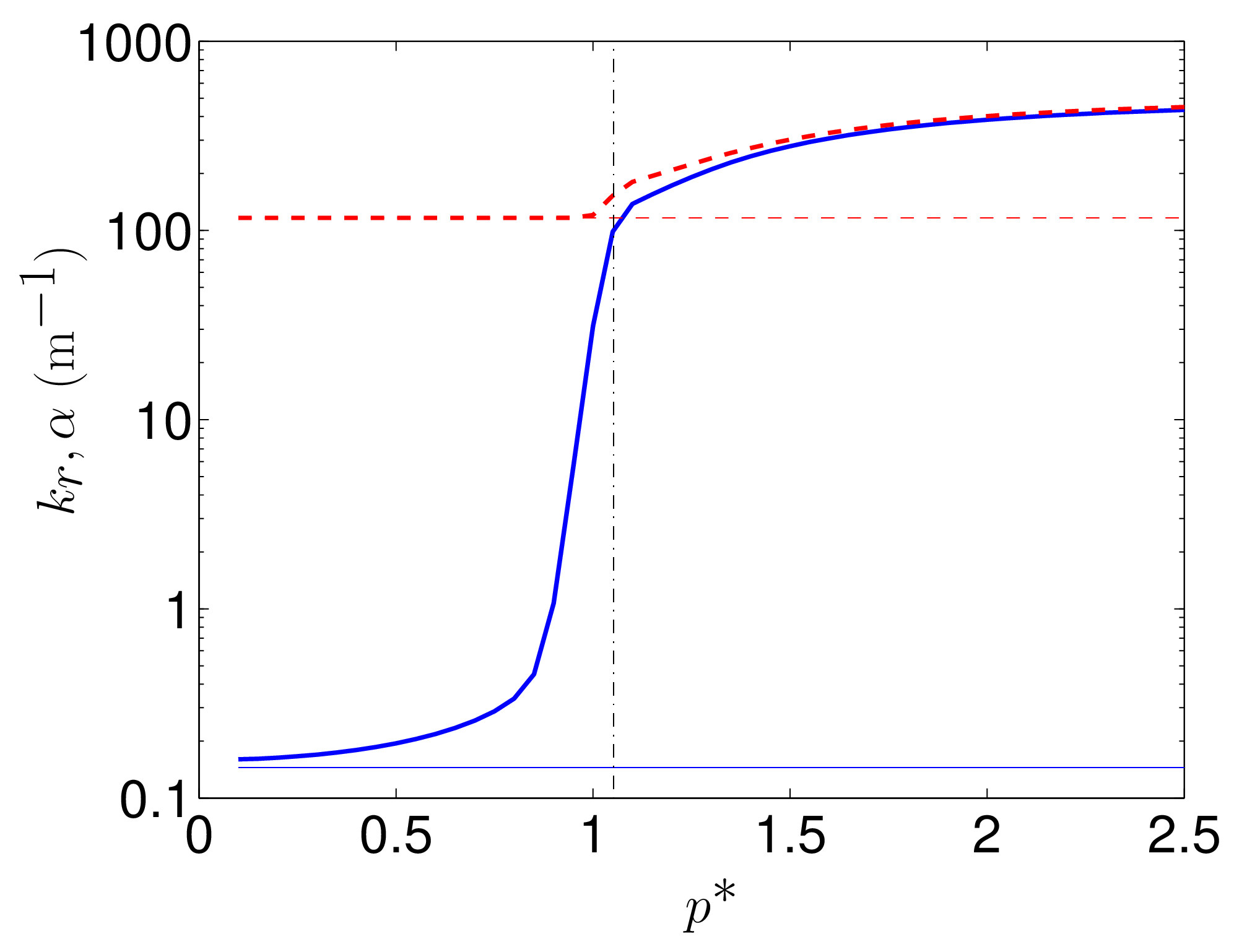}
  \caption{ Real part (dashed) and imaginary part (solid) of the
    wave number $k$. The thin horizontal lines are predictions from
    linear theory (\ref{reldispers}) and the thick lines are results
    calculated from Eqs.~(\ref{defk2i}),~(\ref{defk2r}). The vertical
    dash-dotted line represents the Blake threshold.}
  \label{figalpha}  
\end{figure}

Figure~\ref{figalpha} displays the attenuation coefficient $\alpha$
calculated by following this procedure (thick solid line), for
5~$\mu$m bubbles, and a typical \cite{burdin99} void fraction
$\beta_0=5\times 10^{-5}$, as a function of the acoustic pressure
$|P|$. The attenuation coefficient rises abruptly for acoustic
pressures just above the Blake threshold, as do $\pith$ and $\piv$,
and becomes about 4 orders of magnitude larger than its linear value
[thin solid line, calculated from Eq.~(\ref{reldispers})]. This
demonstrates that a cloud of inertial cavitation bubbles damps out the
incident wave much more drastically than linearly oscillating bubbles.
Moreover, contrarily to the linear prediction, the attenuation
coefficient increases with the wave peak-amplitude. Thus, increasing
the source vibration amplitude does not necessarily produce a more
extended bubble field since increasing the acoustic pressure also
increases the attenuation. This self-saturation phenomenon is
well-known in cavitation experiments~\cite{camposdubus2005}, and will
be demonstrated in the simulations of the next section.

The real part of the wave number is also displayed in
Fig.~\ref{figalpha} (thick dashed line), and the constant linear value
predicted by (\ref{reldispers}) is recalled (thin dashed line) for
comparison. It is interesting to note that, above the Blake threshold,
$k_r$ closely follows $\alpha$. This comes from the fact that the
imaginary part of $k^2$, calculated from the dissipation functions by
Eq. (\ref{defk2i}), is much larger in absolute value than its real
part (\ref{defk2r}). This can be seen by writing the complex
wave number as:
\begin{equation}
  \label{k2expcomp}
  k^2 = K^2 \exp\left[i\left( \epsilon-\pi/2\right) \right],
\end{equation}
where $\epsilon$ is a small number, since $\Im\left(k^2 \right)$ is
negative and large. The wave number $k$  therefore reads:
\begin{equation}
  \label{kexpcomp}
  k = K\exp\left[i\left(\epsilon/2-\pi/4 \right) \right],
\end{equation}
and is therefore almost equal to $K(1-i)/\sqrt{2}$, so that we indeed
have $k_r\simeq\alpha$.

The ratio $\alpha/k_r$ has a strong physical sense. The attenuation of
the wave over one wavelength $\lambda$ is $\exp(-\alpha\lambda) =
\exp(-2\pi\alpha/k_r)$. Thus, if as in the present case $\alpha$ is of
the same order of magnitude as $k_r$, the attenuation of the wave over
one wavelength is of the order of $\exp(-2\pi)\simeq 0.002$. This
means that as soon as the imaginary part of $k^2$ is much larger than
its real part, attenuation will play a dominant role whatever the
precise value of its real value. This is why the precise choice of
$\Re\left(k^2 \right)$ is of minor importance, and Eq. (\ref{defk2r})
is a good compromise. 



\section{Results}
\label{secresults}
\subsection{1D wave profiles}
We consider a tube of length $L$ filled with water, bounded on the
left by a piston which oscillating displacement reads:
\begin{equation}
  \label{defU}
  U(t) = U_0 \cos{\omega t}
\end{equation}
and on the right by an infinitely soft boundary, imposing a zero
acoustic pressure. This arbitrary boundary condition was chosen so
that a standing wave should be obtained in the absence of bubbles. It
can be easily changed to different and more complex conditions, as will
be exemplified in the companion paper.
%

We consider 5~$\mu$m air bubbles. This choice is partially justified
by experimental measurements of bubble size distributions at low
frequency \cite{burdin99,mettin2005}.  In order to solve
(\ref{helmholtzNL}) along with Eqs.~(\ref{defk2i})-(\ref{defk2r}), the
bubble density $N$ must be known. For now, we consider that bubbles are
only present in the zones where the acoustic pressure amplitude is
above the Blake threshold Eq.~(\ref{blake}), and with a uniform density:
\begin{equation}
  \label{choixN}
  N = \left\{
    \begin{array}{lll}
      N_0 & \text{if } |P| > P_B \\
      0 & \text{if } |P| < P_B 
    \end{array}
 \right.
\end{equation}
The nonlinear Helmholtz equation along with (\ref{choixN}) and the
above boundary conditions is solved using the commercial COMSOL
software, and a mesh convergence was performed.

\begin{figure}[h!t]
  \centering
  \includegraphics[width=\linewidth]{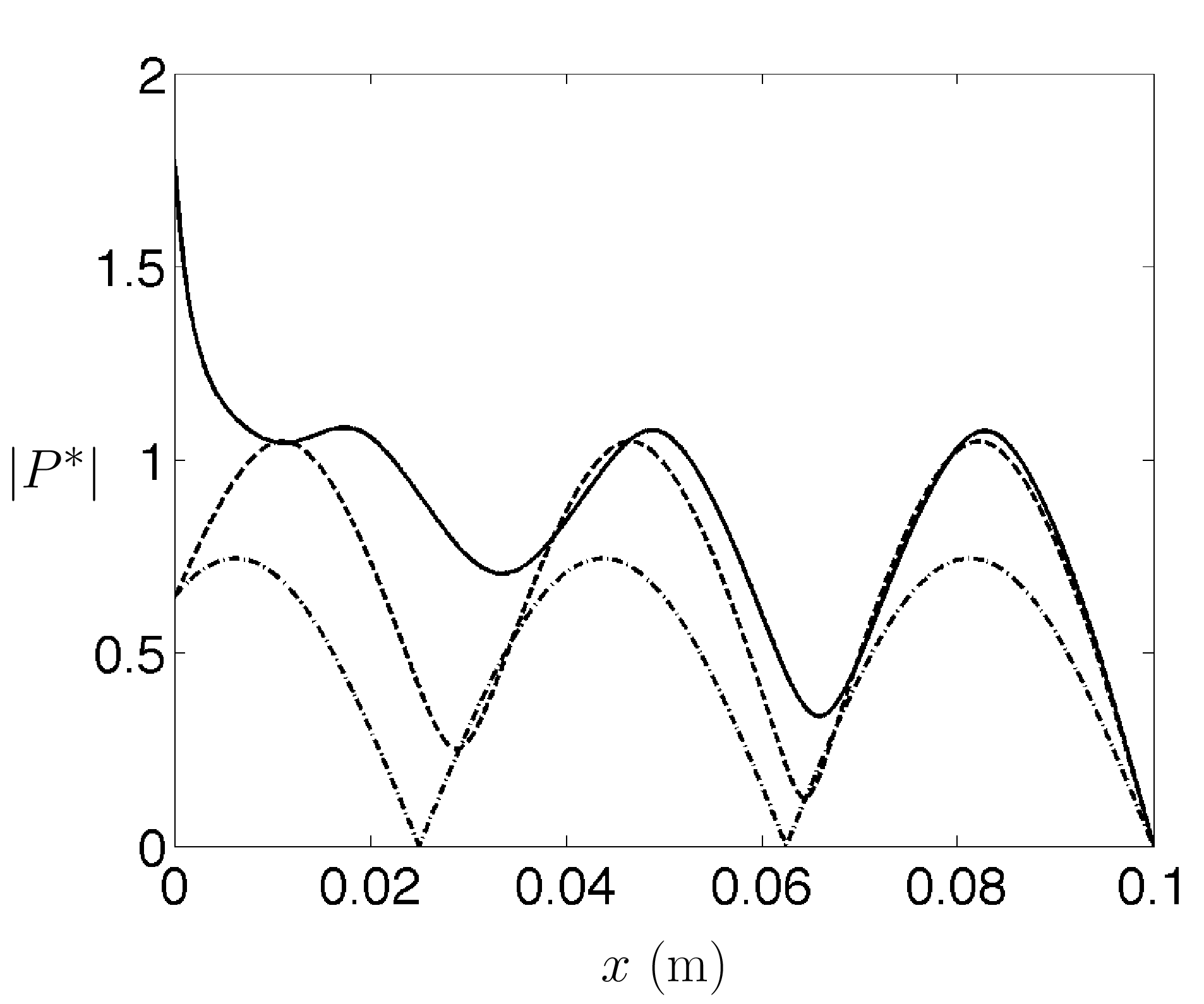}
  \caption{Peak value of the dimensionless pressure field, calculated by
    solving numerically Eq.~(\ref{helmholtzNL}) for various emitter
    displacement amplitudes. Solid line: $U_0=5$~$\mu$m; dashed line:
    $U_0=0.5$~$\mu$m; dash-dotted line: $U_0=0.2$~$\mu$m.}
  \label{figprofpa}
\end{figure}

Figure~\ref{figprofpa} displays the profiles of the peak acoustic
pressure $|\paet|=P/p_0$ obtained for various amplitude of the source. For
the smallest amplitude of the emitter $U_0=0.2$~$\mu$m, we recover a
standing wave profiles in the pure liquid (dash-dotted line). For a
slightly larger vibration of the emitter $U_0=0.5$~$\mu$m (dashed
line), the acoustic pressure at the antinodes is just above the Blake
threshold, so that the bubbles present here start to dissipate some
energy. This yields nonzero acoustic pressures at the nodes, but the
profile remains globally similar to a linear standing wave profile.
When the amplitude of the source is much larger ($U_0=5$~$\mu$m, solid
line), the wave profile completely changes, and is drastically
attenuated in a zone of about 1~cm width near the emitter. This is due
to the fact that the acoustic pressure near the emitter is larger than
the Blake threshold, so that the bubbles present in this zone
dissipate a lot energy. The remaining part of the profile is similar
to a damped linear standing wave.

\begin{figure}[h!t]
  \centering
  \includegraphics[width=\linewidth]{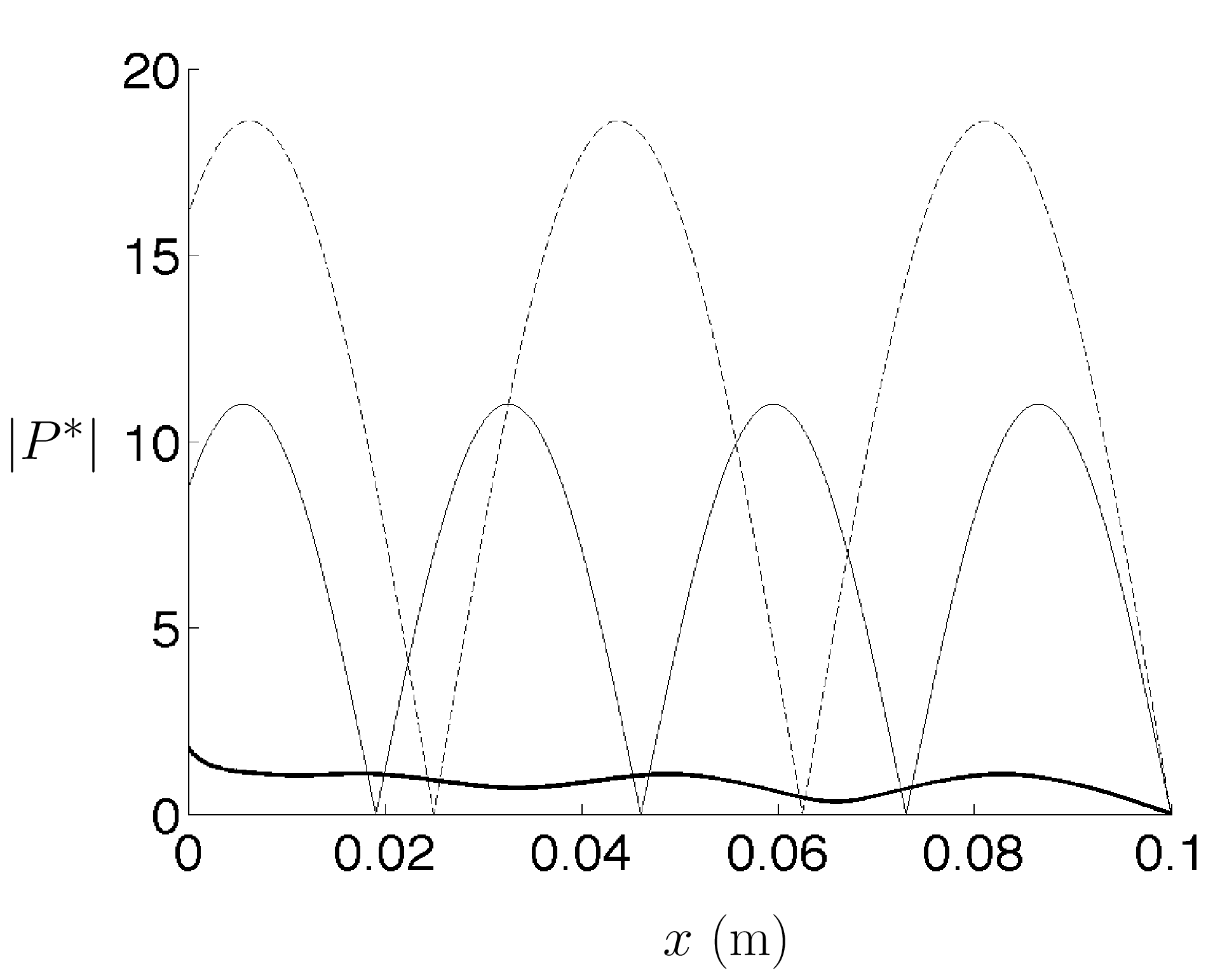}
  \caption{Wave profiles for an amplitude of the emitter of 5~$\mu$m.
    Thick solid curve: predicted by the present model (same as the
    thick solid curve of Fig.~\ref{figprofpa}); thin solid curve:
    obtained by the linear dispersion relation Eq.~(\ref{reldispers});
    thin dashed curve: obtained in the pure liquid.}
  \label{figcomparlin}
\end{figure}

In order to emphasize the importance of the nonlinear energy
dissipation accounted for by our model, we present in
Fig.~\ref{figcomparlin} a comparison of the upper profile of
Fig.~\ref{figprofpa} ($U_0=5$~$\mu$m, thick solid line), to the
profile that would be obtained either by using the linear relation
dispersion~(\ref{reldispers}) with the same bubble density (thin solid
line), or in the pure liquid (thin dashed line). The important
conclusion is that the two linear models predict unrealistic huge
values of the acoustic pressure, while our model yields commonly
measured amplitudes at 20 kHz (typically 1.5-3 bar \cite{mettin2005}).

\subsection{Standing and traveling waves}
The phase $\theta$ between the pressure field and the pressure
gradient allows to determine whether the wave is traveling or
standing. For a purely traveling wave (typically 
$p(x,t)\sim e^{i\left(\omega t - k x \right)}$%
), pressure and
pressure gradient are in phase quadrature, so that $|\sin\theta|=1$.
Conversely, for a purely standing wave (typically
$p(x,t)\sim \cos( k x)e^{i\omega t}$%
), pressure and pressure gradient are in phase or in phase opposition,
so that $|\sin\theta|=0$ in the latter case \cite{morseacous}. Thus,
the quantity $\sin^2\theta$ can be used as a measurement of the
traveling character of the wave.

\begin{figure}[h!t]
  \centering
  \includegraphics[width=\linewidth]{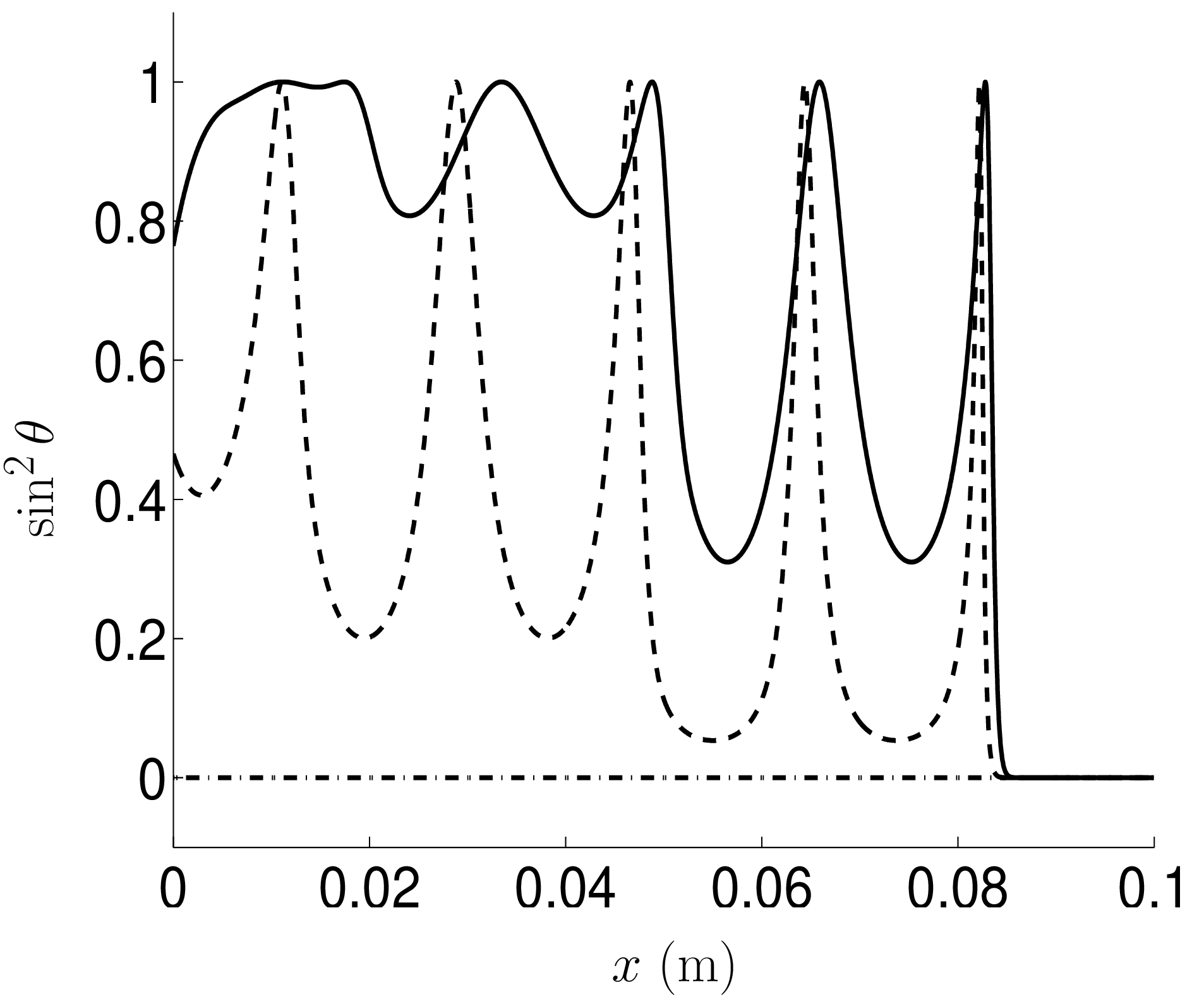}
  \caption{Phase between pressure and pressure gradient in the same
    conditions as Fig.~\ref{figprofpa}. The line-styles are the same as
    for Fig.~\ref{figprofpa}}.
  \label{figphase} 
\end{figure}

In the configuration studied here, where the domain is closed with
perfectly reflecting boundaries, linear acoustics without dissipation
would predict a perfect standing wave.  However, if there is
attenuation in the medium, a traveling wave component appears, because
the reflected wave is of lower amplitude than the incident wave. This
can be checked on Fig.~\ref{figphase}, where $\sin^2\theta$ is
displayed for the same simulation conditions as Fig.~\ref{figprofpa}.
It is seen that for low driving amplitudes ($U_0=$ 0.2 $\mu$m,
dash-dotted line), $\sin^2\theta$ is $0$ everywhere, so that we have
an almost perfect standing wave (which was clearly visible on
Fig.~\ref{figprofpa}). But for higher emitter amplitude ($U_0=$ 0.5
$\mu$m, dashed line), $\sin^2\theta$ starts to increase everywhere in
the medium, especially near the pressure antinodes, and for $U_0=$ 5
$\mu$m (solid line), $\sin^2\theta$ progressively increases toward 1
in a large part of the medium.

\begin{figure}[h!t]
  \centering
  \includegraphics[width=\linewidth]{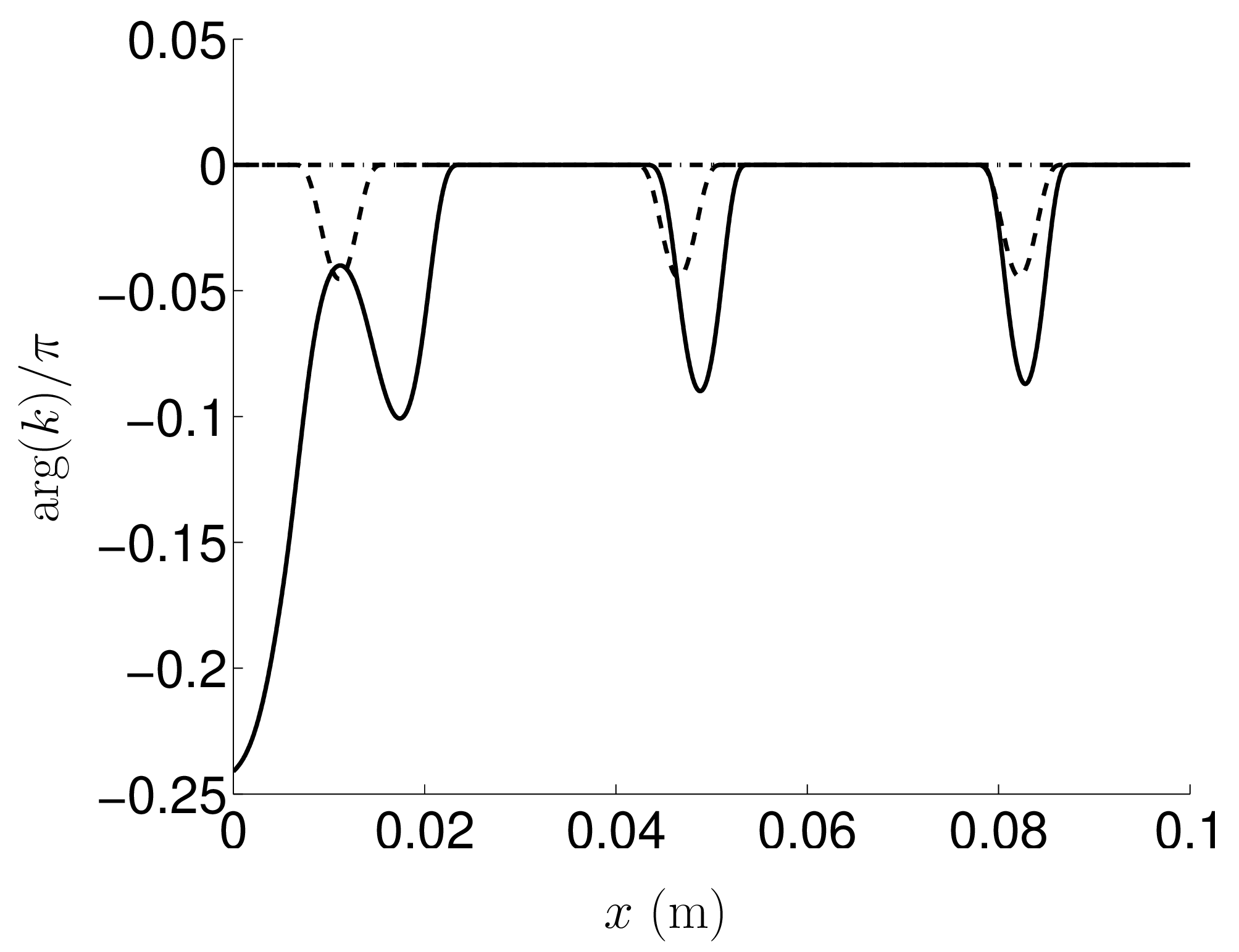}
  \caption{Phase of the complex wave number $k$ divided by $\pi$, in
    the same conditions as Fig.~\ref{figprofpa}, for $U_0=5~\mu$m
    (solid line) $U_0=0.5~\mu$m (dashed line), and $U_0=0.2~\mu$m
    (dashed-dotted line). For the largest amplitude, the wave number
    near the emitter is seen to approach $-\pi/4$, as expected from
    Eq.~(\ref{kexpcomp}).}
  \label{figargk} 
\end{figure}

Finally, Fig.~\ref{figargk} confirms that, as shown above [see
Eq.~(\ref{kexpcomp})], the phase of the complex wave number $k$ is
close to $-\pi/4$ in zones where the bubbles oscillate inertially. The
wave number $k$ is thus proportional to $1-i$, which means that the
attenuation factor $\alpha$ and the real part $k_r$ of the wave number
are of the same orders of magnitude.

\section{Conclusion}
\label{secconclusion} 
Inertial bubbles dissipate much more energy than a linearly oscillating
bubble, both by thermal diffusion in the gas and viscous dissipation
in the liquid, the latter mechanism being dominant for bubble ambient
radii lower than 10 $\mu$m. The wave attenuation in an inertial
cavitation field is therefore much larger than the value predicted by
the classical linear dispersion relation (by typically 4 orders of
magnitude).  Although the latter conclusion is qualitatively
intuitive, to our knowledge, no quantitative estimation has ever been
reported.


Under the assumption that the bubbles are mainly excited by the first
harmonic content of the acoustic field, the latter fulfills
approximately a nonlinear Helmholtz equation. The imaginary part of
the squared wave number is estimated rigorously from the energy
dissipated by a single bubble, which can be easily calculated by
solving a bubble dynamic equation. The real part is still arbitrarily
estimated from the linear theory, but this arbitrary choice was shown
to be of low importance, owing to the the huge value of the imaginary
part. This has the importance consequence that in bubbly zones, the
attenuation factor is of the same order of magnitude as the real part
of the wave number, which results in a strong attenuation of the wave.

The model has been solved in a typical 1D-domain, and yields as
expected a strongly attenuated wave profile near the emitter for high
amplitude vibrations of the latter. The amplitude of the calculated
acoustic pressure fields are realistic, contrarily to linear theory.
This strong attenuation yields in turn a traveling component in the
wave, where purely standing waves would be expected in a non dissipative
medium enclosed by perfectly reflecting boundaries.

It is interesting to note that following the present results,
attenuation, and therefore wave structures, are mainly governed by viscous
dissipation involved in the bubble radial motion, the thermal effects in the
bubble playing a minor role. This conjecture might be checked
experimentally by measuring the wave attenuation for solutions of
different viscosities and with different dissolved gas. 

\olchange{The choice of the incompressible Rayleigh-Plesset equation
  to model the bubble dynamics may be questioned. Although this is the
  original formulation of the Caflish model, the compressibility of
  the liquid produces sound scattering, and contributes therefore to
  the attenuation of waves in bubbly liquids, as is well-known in the
  linear case \cite{foldy,commprosper}. One may therefore replace
  Eq.~(\ref{rayleigh}) for example by a Keller equation
  \cite{kellermiksis,prosperlezzi1,doinikov2005review}, and
  reformulate the energy equation (\ref{eqKbubblemoy}) to exhibit an
  additional contribution of radiation $\Pi_a$ in its right-hand-side.
  This would in turn add a contribution in Eq.~(\ref{defk2i}), and
  produce more wave attenuation.  However, the procedure is not
  straightforward, and the energetical interpretation in this case is
  less easy. One of the reasons for that is that compressible bubble
  dynamics equations are not exact solutions of the basic physical
  principles \cite{prosperlezzi1,doinikov2005review}, but only first
  terms of expansions in the parameter $\dot{R}/c_l$. It is also
  expected that sound scattering also modifies the real part of $k^2$,
  which again raises the issue of a correct expression for the latter.
  However, it may be conjectured that, in the low frequency range
  studied here, the power loss by sound scattering is much lower than
  the one produced by viscous dissipation, because, as for thermal
  effects, sound radiation occurs mainly in the vicinity of the
  collapse. Thus we expect that the model in its present form catches
  the main dissipation phenomenon and that the values proposed for
  $\Im(k^2)$ is a good estimation. This will be examined in more
  details in future work.}

The occurrence of traveling waves, aside of the issue of the Bjerknes
forces examined in the companion paper, may also have fundamental
consequences on the final stage of the bubbles collapse. Indeed, it
has been shown recently that bubbles in traveling waves are more
exposed to shape instabilities and can undergo jetting, which reduces
the final collapse temperature \cite{calvisiszeri2007}, compared to a
spherically collapsing bubble. This would therefore influence the
estimation of the heat lost by a single bubble, but the spherical
collapse model used in the present study yields an upper value.

Besides, measurements of the acoustic field in conical structures has
revealed the presence of a time-independent mean pressure field, which
amplitude may be comparable with the first-harmonic part
\cite{camposdubus2005}. Our model does not catch this feature, and
there is yet no correct theoretical description of this phenomenon. We
emphasize however that our derivation of the imaginary part of the
wave number is valid even in this case, since our decomposition of the
field Eq.~(\ref{pdecomp}) accounts a priori for the presence of such a
mean field. This suggests that the present model could be supplemented
by a specific equation describing this mean pressure field, which
remains to be determined.
 
To conclude, we believe that the present model opens the way to more
realistic simulations of the coupled evolution of the cavitation field
and the acoustic field.  The nonlinear Helmholtz equation is
relatively easy to solve and constitutes a viable solution halfway
between a fully nonlinear simulation of the Caflish equations, which
requires painful, if not intractable, temporal integration, and a
fully linearized model which, as shown above, yields unrealistic
acoustic pressure values. The companion paper will address the
calculation of the Bjerknes forces in the acoustic fields calculated
with the present model, and the resulting bubble structures predicted
in more complex 2D configurations.


\section{Acknowledgments}
The author acknowledges the support of the French Agence Nationale de
la Recherche (ANR), under grant SONONUCLICE (ANR-09-BLAN-0040-02)
''Contr\^ole par ultrasons de la nucl\'eation de la glace pour
l'optimisation des proc\'ed\'es de cong\'elation et de
lyophilisation''. Besides the author would like to thank Nicolas Huc
of COMSOL France for his help in solving convergence issues.

\vfill
\pagebreak

\bibliographystyle{elsart-num-usson}


\pagebreak
\listoffigures

\end{document}